\documentclass[conference]{IEEEtran}
\IEEEoverridecommandlockouts

\usepackage{cite}
\usepackage{amsmath,amssymb,amsfonts}
\usepackage{algorithmic}
\usepackage{graphicx}
\usepackage{textcomp}
\usepackage{xcolor}
\usepackage{tabularx}

\usepackage{cite}
\usepackage{graphicx}
\usepackage{amsmath,amssymb}
\usepackage{booktabs}
\usepackage{url}
\usepackage{xcolor}
\usepackage{cite}
\usepackage{amsmath,amssymb,amsfonts}
\usepackage{algorithmic}
\usepackage{graphicx}
\usepackage{textcomp}
\usepackage{xcolor}
\usepackage{hyperref}
\usepackage{booktabs}
\usepackage{multirow}
\usepackage{array}
\usepackage[most]{tcolorbox}
\def\BibTeX{{\rm B\kern-.05em{\sc i\kern-.025em b}\kern-.08em
    T\kern-.1667em\lower.7ex\hbox{E}\kern-.125emX}}

\usepackage{tikz}
\usepackage{textcomp}

\newcommand\copyrighttext{%
  \footnotesize \textcopyright 2026 IEEE. Personal use of this material is permitted.  Permission from IEEE must be obtained for all other uses, in any current or future media, including reprinting/republishing this material for advertising or promotional purposes, creating new collective works, for resale or redistribution to servers or lists, or reuse of any copyrighted component of this work in other works. 

  Accepted as paper at SAGAI-ICSA2026.}
\newcommand{\copyrightnotice}{%
\begin{tikzpicture}[remember picture,overlay]
\node[anchor=south,yshift=10pt] at (current page.south) {\fbox{\parbox{\dimexpr\textwidth-\fboxsep-\fboxrule\relax}{\copyrighttext}}};
\end{tikzpicture}%
}

\begin{document}

\title{Designing Adaptive Digital Nudging Systems with LLM-Driven Reasoning}

\author{\IEEEauthorblockN{Tiziano Santilli}
\IEEEauthorblockA{\textit{Mærsk Mc-Kinney Møller Instituttet} \\
\textit{Syddansk Universitet}\\
Odense, Denmark}
\and
\IEEEauthorblockN{Mina Alipour}
\IEEEauthorblockA{\textit{Mærsk Mc-Kinney Møller Instituttet} \\
\textit{Syddansk Universitet}\\
Odense, Denmark}
\and
\IEEEauthorblockN{Mahyar Tourchi Moghaddam}
\IEEEauthorblockA{\textit{Mærsk Mc-Kinney Møller Instituttet}} 
\textit{Syddansk Universitet}\\
Odense, Denmark}

\maketitle
\copyrightnotice
\begin{abstract}
Digital nudging systems lack architectural guidance for translating behavioral science into software design. While research identifies nudge strategies and quality attributes, existing architectures fail to integrate multi-dimensional user modeling with ethical compliance as architectural concerns. We present an architecture that uses behavioral theory through explicit architectural decisions, treating ethics and fairness as structural guardrails rather than implementation details. A literature review synthesized 68 nudging strategies, 11 quality attributes, and 3 user profiling dimensions into architectural requirements. The architecture implements sequential processing layers with cross-cutting evaluation modules enforcing regulatory compliance. Validation with 13 software architects confirmed requirements satisfaction and domain transferability. An LLM-powered proof-of-concept in residential energy sustainability demonstrated feasibility through evaluation with 15 users, achieving high perceived intervention quality and measurable positive emotional impact. This work bridges behavioral science and software architecture by providing reusable patterns for adaptive systems that balance effectiveness with ethical constraints.
\end{abstract}

\begin{IEEEkeywords}
Software Architecture, Digital Nudging, Large Language Models, Behavioral Theory, Ethical AI, Quality Attributes, Architectural Decisions
\end{IEEEkeywords}

\section{Introduction}

The role of software systems is evolving from passive tools to active participants in human decision-making. This shift is exemplified by the Internet of Behaviors (IoB), which extends the Internet of Things by connecting user data to behavioral insights~\cite{iob1,iob2}. Digital nudging represents a key mechanism for this interaction, using behavioral science to guide users toward beneficial choices in domains like energy conservation~\cite{energy} and health improvement~\cite{health}. However, designing adaptive nudging systems presents fundamental architectural challenges that current practice fails to address systematically.

Digital nudging presents three architectural challenges that cannot be resolved through implementation alone:

{\em i)} Multi-dimensional runtime adaptation. Effective nudging requires simultaneous consideration of cognitive mode (System 1 vs. System 2 thinking), behavioral stage (pre-contemplation through maintenance), attention capacity, and quality attributes~\cite{P18,P22}. This creates the modifiability-performance trade-off~\cite{bass2021}: tight coupling enables coordinated responses but creates brittleness; loose coupling improves maintainability but complicates adaptation. The architectural question becomes: how do we structure components such that changes in one dimension (e.g., attention drops) trigger coordinated adaptations across multiple subsystems (UI simplification, strategy adjustment, message adaptation) without tight coupling?

{\em ii)} Non-negotiable ethical constraints as architectural guardrails. The EU AI Act, GDPR Article 22, and Digital Services Act impose hard constraints on automated decision-making systems~\cite{aiact,gdpr,dsa}. These are not optional features but mandatory architectural properties that must hold across all execution paths. Treating ethics as an afterthought leads to architectures where manipulative strategies can slip through during edge cases. The architectural challenge is: how do we enforce ethical boundaries structurally, such that no component can generate non-compliant nudges regardless of user state or system evolution?

{\em iii)} Quality attribute conflicts requiring explicit prioritization. Literature reveals inherent tensions: transparency increases trust but may trigger reactance~\cite{P10,P12}; personalization improves effectiveness but raises fairness concerns~\cite{P25}; simplification aids low-attention users but limits information for analytical thinkers~\cite{P18,P22}. These conflicts demand architectural tactics, like explicit design decisions that prioritize certain qualities at the expense of others, with documented rationale and trade-offs.

Current practice treats user adaptation as an implementation concern rather than an architectural requirement, resulting in systems that cannot guarantee consistent ethical operation or adapt systematically~\cite{Arch1,Arch2}. Software architects lack systematic guidance for translating behavioral science principles into concrete architectural decisions, Section~\ref{gap}.

To address these challenges, we must understand the theoretical foundation. Behavioral theory, originating with Watson's behaviorism~\cite{watson1913psychology} and advanced by Skinner's operant conditioning~\cite{skinner1938behavior,skinner1953science}, posits that human actions result from conditioned responses to environmental stimuli. Learning is defined as measurable behavioral change through experience. Despite this theoretical background, the software engineering community lacks systematic approaches to address these concepts architecturally.

This paper addresses these gaps through four contributions: {\em i)} a literature review synthesizing nudging strategies, quality attributes, and user profiling dimensions into architectural requirements; {\em ii)} an architecture treating behavioral dimensions and ethical compliance as first-class architectural concerns; {\em iii)} validation with software architects confirming requirements satisfaction and transferability; {\em iv)} a proof-of-concept implementation using large language models (LLMs) for cognitive reasoning, with empirical user evaluation.

We address three research questions:
\begin{itemize}
    \item \textbf{RQ1:} What are the key characteristics, quality attributes, and metrics for digital nudging?
    \item \textbf{RQ2:} To what extent can a software architecture integrate nudging characteristics with intervention logic while preserving modularity and ethical compliance?
    \item \textbf{RQ3:} To what extent does the instantiated architecture demonstrate feasibility and generate effective adaptive interventions?
\end{itemize}

\begin{figure}[h]
\centering
\includegraphics[width=0.45\textwidth]{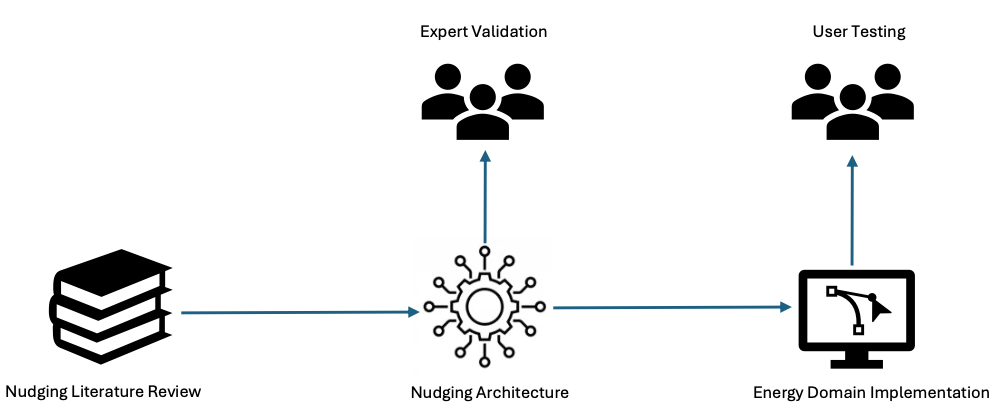}
\caption{Research Methodology Overview}
\label{fig:method}
\end{figure}

For the methodology we employ a four-phase approach Figure~\ref{fig:method}: {\em i)} Literature Review synthesizing behavioral characteristics into requirements; {\em ii)} Architectural Design translating requirements into components with explicit decision rationale; {\em iii)} Architect Evaluation validating requirements satisfaction and transferability; {\em iv)} Implementation \& User Evaluation demonstrating feasibility through proof-of-concept.

The paper proceeds as follows: Section~\ref{sec:slr} reports literature review and gap analysis; Section~\ref{sec:arch} presents the architecture with architectural decisions and validation; Section~\ref{sec:prototype} describes implementation and user evaluation; Section~\ref{sec:threats} discusses validity threats; Section~\ref{sec:conc} concludes with future work. \textbf{Supplemental materials} available at this link\footnote{https://github.com/tiziasan/Adaptive-Digital-Nudging-System}

\section{Nudging Literature Review}
\label{sec:slr}

To establish an understanding of digital nudging characteristics, we conducted a literature review.

\subsection{Review Process}

We queried four academic databases (IEEE Xplore, Scopus, ACM Digital Library, Web of Science) using the search string: \texttt{((nudge* OR nudging) AND ("quality attribute*" OR metric* OR framework OR characteristic*))}. The search yielded 707 papers. After removing 308 duplicates and applying inclusion/exclusion criteria, full-text review resulted in 21 primary studies. Inclusion/exclusion criteria list and literature review flow are available in the \textbf{supplemental materials}.
\subsection{Results}

Our review identified three key contributions:

{\em i)} Nudging Strategies (N = 68). We extracted a taxonomy spanning information provision (anchoring, framing, simplification), social influence (social norms, peer support), defaults and positioning, feedback mechanisms, and timing strategies. The complete taxonomy is available in \textbf{supplemental materials}.

{\em ii)} Quality Attributes (N = 11). Literature identifies attributes determining effectiveness and ethicality: acceptability, awareness, effectiveness, fairness, healthfulness, helpfulness, intrusiveness, motivation, transparency, trustworthiness, and user impact. Table~\ref{tab:quality} presents definitions.

\begin{table}[t]
\caption{Quality Attributes for Digital Nudging}
\label{tab:quality}
\centering
\footnotesize
\begin{tabular}{p{2cm}p{5cm}p{0.7cm}}
\toprule
\textbf{Attribute} & \textbf{Definition} & \textbf{Ref} \\
\midrule
Acceptability & Degree to which users perceive a nudge as appropriate or desirable, considering social, cultural, moral standards & \cite{P10} \\
Awareness & Degree of user attention to the nudge. Tension exists: high transparency may trigger reactance, subliminal approaches raise ethical concerns & \cite{P10} \\
Effectiveness & Degree to which a nudge achieves its intended behavioral outcome & \cite{P11} \\
Fairness & Ensures systems address algorithm/data biases and cognitive/perceptual biases, providing equitable access. Includes GDPR, AI Act, DSA compliance & \cite{P25} \\
Healthfulness & Degree to which users perceive an item or choice to impact their health positively & \cite{P10} \\
Helpfulness & Degree to which users perceive nudges as assisting decision-making & \cite{P10} \\
Intrusiveness & Low-intrusive nudges subtly guide decisions; high-intrusive nudges strongly influence decisions & \cite{P11} \\
Motivation & Degree to which users are motivated to make choices aligned with the nudge's goal.  & \cite{P10} \\
Transparency & Explainability frameworks for DSA/AI Act requirements while maintaining effectiveness & \cite{P10,P13,P25} \\
Trustworthiness & Degree to which users trust the nudge and information provided & \cite{P10} \\
User Impact & Effect of nudge on user's goals, preferences, or well-being & \cite{P11} \\
\bottomrule
\end{tabular}
\end{table}

{\em iii)} User Profiling Dimensions (N = 3). Literature identifies three behavioral dimensions:

\textit{Cognitive Mode:} Nudges target either System 1 (fast, intuitive) or System 2 (slow, analytical) thinking~\cite{P18,P22}. System 1 nudges leverage automatic processing; System 2 nudges support deliberate reasoning.

\textit{Behavioral Stage:} The Transtheoretical Model recognizes five stages~\cite{P18}: pre-contemplation (unaware), contemplation (considering change), preparation (planning action), action (actively modifying behavior), maintenance (sustaining change). Each stage requires different intervention strategies.

\textit{Attention Capacity:} Users make choices attentively or inattentively based on cognitive load~\cite{P22,P24}. Low attention requires simple nudges; high attention enables information-rich strategies.

\begin{tcolorbox}[colback=gray!10,boxrule=0.5pt,title=RQ1 Answer]
Our literature review identified 68 nudging strategies, 11 quality attributes, and 3 user profiling dimensions. These constitute requirements for digital nudging systems that align interventions with user states.
\end{tcolorbox}

\subsection{Gap Analysis}
\label{gap}

{\em i)} Digital Nudging Architectures. Karlsen and Andersen present a smart nudging system with user profile learner and nudge design components~\cite{Arch1,Arch2}, but assess only interests, capabilities, and activity history, omitting cognitive mode, attention state, and quality attributes as architectural dimensions. Their system represents closest prior work but lacks systematic integration of behavioral theory dimensions.

{\em ii)} Behavioral Theories in Software. Grisiute and Raubal explore spatial nudging converging persuasive technologies with spatial design~\cite{P18}, but do not provide architectural guidance. Lofgren and Nordblom present theoretical framework of decision-making explaining nudging mechanisms~\cite{P22}, but focus on psychological models rather than software architecture.

{\em iii)} Ethical AI Architectures. Milano et al. discuss recommender systems' ethical challenges~\cite{P25}, emphasizing fairness as critical concern. Fabbri examines self-determination through explanation for DSA transparency requirements~\cite{P13}. Our architecture operationalizes these ethical concerns through structural enforcement rather than post-hoc auditing.

{\em iv)} Quality-Driven Architectures. Bass et al.'s quality attribute scenarios~\cite{bass2021} provide methodological foundation for our approach. We extend this by showing how behavioral characteristics  function as quality attributes driving architectural decisions, alongside traditional attributes.

\section{Digital Nudging Architecture}
\label{sec:arch}

We translate literature findings into an architecture addressing the identified gaps. This section presents requirements with traceability, architectural design with explicit decisions, and validation results.

\subsection{Requirements Traceability}

Table~\ref{tab:requirements} presents functional requirements (FR) and non-functional requirements (NFR) derived from literature.

\begin{table*}[t]
\caption{Architectural Requirements with Literature Traceability}
\label{tab:requirements}
\centering
\footnotesize
\begin{tabular}{p{0.6cm}p{4cm}p{3.5cm}p{6cm}}
\toprule
\textbf{ID} & \textbf{Requirement} & \textbf{Component} & \textbf{Derivation \& References} \\
\midrule
\multicolumn{4}{c}{\textit{Functional Requirements}} \\
\midrule
FR1 & Capture context and behavioral data & Context Tracker, Behavioral Collector & Context-awareness~\cite{P14}, behavioral signals~\cite{P9,P18} \\
FR2 & Determine cognitive mode & Cognitive Mode & Dual-process theory~\cite{P18,P22} \\
FR3 & Detect behavioral stage & Behavioral Stage & Transtheoretical Model~\cite{P1,P18} \\
FR4 & Estimate attention capacity & Attention Capacity & Cognitive load theory~\cite{P22,P24} \\
FR5 & Generate diverse nudges & Nudge Generator & Strategy taxonomy~\cite{P1,P7,P11} \\
FR6 & Optimize strategy selection & Strategy Optimizer & Architectural necessity for multi-dimensional constraint satisfaction \\
FR7 & Provide transparency & Explainability & Regulatory requirements~\cite{P10,P12,P13} \\
FR8 & Capture user feedback & User Feedback & Adaptive refinement~\cite{P9,P15} \\
FR9 & Track outcomes & Outcome Tracker & Effectiveness measurement~\cite{P11} \\
\midrule
\multicolumn{4}{c}{\textit{Non-Functional Requirements}} \\
\midrule
NFR1 & Ethics compliance & Ethics Compliance & GDPR/AI Act/DSA mandates~\cite{aiact,gdpr,dsa,P12,P20} \\
NFR2 & Fairness monitoring & Fairness Monitor & Bias mitigation~\cite{P25} \\
NFR3 & Transparency & Explainability & DSA Article 27, AI Act Article 13~\cite{P10,P13,P25} \\
NFR4 & Adaptivity & Adaptive Dashboard, UI Adaptation & State-dependent effectiveness~\cite{P9,P14,P18} \\
\bottomrule
\end{tabular}
\end{table*}

FR6 Rationale: Strategy Optimizer emerges from architectural necessity, not explicit literature mention. When multiple strategies match user state (e.g., System 1 thinking + pre-contemplation + low attention), we need conflict resolution logic to select the single most appropriate intervention. This component implements constraint satisfaction reasoning.

NFR Selection Rationale: From 11 quality attributes (Table~\ref{tab:quality}), we prioritized 4 as NFRs based on: {\em i)} Regulatory mandates,  ethics Compliance and Fairness address legally required properties under GDPR Article 22, AI Act Article 10, DSA Article 27; {\em ii)} Architectural impact, Transparency and Adaptivity fundamentally shape system structure. The remaining attributes are evaluation metrics measured during user testing (Section~\ref{sec:user}).

\subsection{Architecture Description}

\begin{figure}[h]
\includegraphics[width=0.5\textwidth]{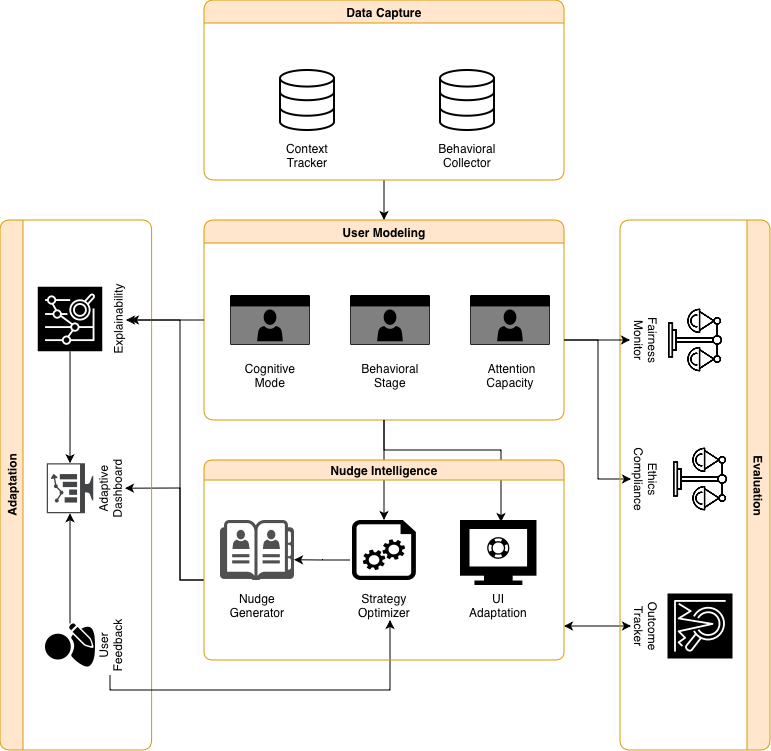}
\caption{Digital Nudging Architecture}
\label{fig:architecture}
\end{figure}

Figure~\ref{fig:architecture} presents our architecture comprising three core processing layers (Data Capture, User Modeling, Nudge Intelligence) with two cross-cutting modules (Adaptation, Evaluation).

{\em i)} Data Capture captures raw signals through Context Tracker (device, location, time-of-day) and Behavioral Collector (clicks, hesitation times, navigation patterns, facial emotions).

{\em ii)} User Modeling transforms signals into multi-dimensional profiles through three parallel components: Cognitive Mode (System 1 vs. System 2), Behavioral Stage (Transtheoretical Model mapping), Attention Capacity (high/medium/low).

{\em iii)} Nudge Intelligence generates personalized interventions through Strategy Optimizer that selects optimal strategy given user profile, Nudge Generator that creates message content, and UI Adaptation that specifies interface parameters.

{\em iv)} Adaptation Module (left side) implements delivery through Explainability generating transparency reports, Adaptive Dashboard rendering dynamic interface, and User Feedback capturing explicit ratings.

{\em v)} Evaluation Module (right side) ensures quality through Fairness Monitor that detects population-level biases, Ethics Compliance that validates regulatory constraints, and Outcome Tracker that logs execution traces.

\subsection{Architectural Decisions and Rationale}

We document key architectural decisions following the template: Decision, Rationale, Trade-offs, and Alternatives Considered in Table \ref{tab:architectural-decisions}.

\begin{table*}[t]
\centering
\footnotesize
\label{tab:architectural-decisions}
\caption{Architectural Decisions and Rationale}
\begin{tabularx}{\textwidth}{|p{0.18\textwidth}|X|}
\hline
\textbf{Decision} & \textbf{Details} \\
\hline
\hline
\multicolumn{2}{|l|}{\textbf{AD1: Sequential Pipeline vs. Event-Driven Architecture}} \\
\hline
Decision & Three sequential layers with per-user session processing rather than event-driven publish-subscribe. \\
\hline
Rationale & Digital nudging requires deterministic reasoning chains where User Modeling completes before Strategy Optimization. Sequential processing ensures causal consistency, reasoning transparency, and provenance tracking. \\
\hline
Trade-offs & \textbf{Gains}—Clear provenance, easier debugging, compliance auditability (GDPR Article 22), simplified testing. \textbf{Costs}—No intra-user parallelization, increased per-user latency. \\
\hline
Alternatives & Event-driven with orchestration (rejected: excessive overhead), blackboard architecture (rejected: non-deterministic control flow), pipes-and-filters (rejected: unnecessary given sequential requirements). \\
\hline
\hline
\multicolumn{2}{|l|}{\textbf{AD2: Side-Mounted Evaluation vs. Embedded Validators}} \\
\hline
Decision & Ethics Compliance as side-mounted module intercepting all Nudge Intelligence outputs. \\
\hline
Rationale & Enforces separation of concerns: nudge generation focuses on effectiveness; ethics audits all outputs. Interceptor pattern ensures no nudge reaches users without validation. Provides architectural evidence for DSA Article 27 compliance. \\
\hline
Trade-offs & \textbf{Gains}—Consistent enforcement, evolvability, independent testability, regulatory evidence. \textbf{Costs}—Additional latency, potential bottleneck. \\
\hline
Alternatives & Embedded validators (rejected: duplication risk), post-hoc filtering (rejected: lacks explicit architectural representation), aspect-oriented programming (rejected: deployment complexity). \\
\hline
\hline
\multicolumn{2}{|l|}{\textbf{AD3: LLM-Driven Reasoning vs. Rule-Based Classification}} \\
\hline
Decision & User Modeling components use LLMs rather than rule-based systems. \\
\hline
Rationale & Behavioral constructs like "System 1 thinking" require contextual interpretation of multiple weak signals~\cite{P19,P22}. LLMs handle ambiguous cases (e.g., high clicks but short duration). Sequential pipeline ensures deterministic data flow despite LLM stochasticity within classifications. \\
\hline
Trade-offs & \textbf{Gains}—Robustness to edge cases, expressiveness, adaptability via prompt engineering. \textbf{Costs}—Controlled non-determinism, explainability challenges, API latency (200-500ms), vendor dependency. \\
\hline
Alternatives & Threshold rules (rejected: lacks semantic understanding), supervised ML classifiers (rejected: insufficient for nuanced behavioral patterns), hybrid approach (future work). \\
\hline
\hline
\multicolumn{2}{|l|}{\textbf{AD4: Backend-Driven UI Adaptation vs. Client-Side Adaptation}} \\
\hline
Decision & Backend generates structured "UI Context" payloads specifying interface parameters sent to frontend. \\
\hline
Rationale & Attention capacity and cognitive mode affect information processing~\cite{P18,P22}. Backend determines \textit{what} adaptations based on User Modeling; frontend determines \textit{how} to render. Implements Strategy Pattern keeping behavioral intelligence server-side. \\
\hline
Trade-offs & \textbf{Gains}—Personalization, reduced cognitive load, accessibility compliance, privacy protection. \textbf{Costs}—Frontend complexity, consistency challenges, increased payload size. \\
\hline
Alternatives & Frontend-driven (rejected: exposes interaction data to browser), progressive disclosure (rejected: orthogonal concern), static templates (rejected: cannot provide fine-grained adaptation). \\
\hline
\end{tabularx}

\end{table*}

\subsection{Architectural Evaluation}
\label{sec:expert}

We evaluated the architecture through two methods: {\em i)} questionnaire assessing requirements satisfaction, {\em ii)} follow-up discussions exploring architectural insights.

 We recruited 13 software architects through professional networks. Selection criteria: {\em i)} experience with quality-attribute-driven projects, {\em ii)} familiarity with adaptation systems. Table~\ref{tab:demographics} presents demographics.

\begin{table}[h]
\caption{Architectural Evaluators Demographics (N=13)}
\label{tab:demographics}
\centering
\footnotesize
\begin{tabular}{llc}
\toprule
\textbf{Characteristic} & \textbf{Category} & \textbf{N (\%)} \\
\midrule
Experience & 0--5 years & 4 (30.8\%) \\
 & 5--10 years & 5 (38.5\%) \\
 & 10--15 years & 1 (7.7\%) \\
 & 15+ years & 3 (23.1\%) \\
\midrule
Domain & Healthcare & 3 (23.1\%) \\
 & Automotive & 4 (30.8\%) \\
 & Consultancy & 3 (23.1\%) \\
 & Other & 3 (23.1\%) \\

\bottomrule
\end{tabular}
\end{table}

 Participants received: {\em i)} architecture diagram, {\em ii)} requirements table, {\em iii)} architectural decisions document. They completed a  questionnaire with 16 closed-ended questions using 5-point Likert scales (1=strongly disagree, 5=strongly agree). Questions assessed whether components  met requirements and questionnaire data available in the \textbf{supplemental materials}.

 Table~\ref{tab:surveyresults} presents results. Mean rating across all items was 4.62 (SD=0.12), indicating strong agreement. UI Adaptation achieved highest rating (M=4.85, SD=0.38); Explainability received lowest (M=4.38, SD=0.65), suggesting this is the most contentious aspect.

\begin{table}[h]
\caption{Structured Questionnaire Results (Likert Scale 1-5)}
\label{tab:surveyresults}
\centering
\footnotesize
\begin{tabular}{p{4.5cm}c}
\toprule
\textbf{Requirement} & \textbf{Mean} \\
\midrule
\multicolumn{2}{c}{\textit{Architectural Quality}} \\
Maintainability \& Evolvability & 4.62 \\
Regulatory Compliance & 4.69 \\
\midrule
\multicolumn{2}{c}{\textit{Functional Requirements}} \\
Context/Behavioral Data (FR1) & 4.54 \\
Cognitive Mode (FR2) & 4.62 \\
Behavioral Stage (FR3) & 4.62 \\
Attention Capacity (FR4) & 4.54 \\
Nudge Generator (FR5) & 4.69 \\
Strategy Optimizer (FR6) & 4.69 \\
Explainability (FR7) & 4.38 \\
User Feedback (FR8) & 4.46 \\
Outcome Tracker (FR9) & 4.54 \\
\midrule
\multicolumn{2}{c}{\textit{Non-Functional Requirements}} \\
Ethics Compliance (NFR1) & 4.54 \\
Fairness Monitor (NFR2) & 4.77 \\
Transparency (NFR3) & 4.69 \\
UI Adaptation (NFR4) & 4.85 \\
\midrule
\multicolumn{2}{c}{\textit{Domain Generalizability}} \\
Highly Transferable & 61.5\% \\
Transferable with modifications & 38.5\% \\
\bottomrule
\end{tabular}
\end{table}

We conducted a discussion with 5 of 13 participants (others declined due to time constraints). Discussions were semi-structured around: (1) architectural strengths/weaknesses, (2) missing concerns, (3) transferability.

 Two participants highlighted separation between behavioral reasoning and ethical enforcement:

\textit{"I like that ethics is not inside the nudge generator. That's a recipe for unexpected behaviors. Having it as a separate point makes it testable and auditable."} , P3.

 Three participants raised concerns:

\textit{"Using LLMs for cognitive mode worries me. How do you regression test that? If GPT gets updated and starts classifying differently, your system behavior changes. You need to carefully manage the LLM non determinism"}, P2.

All 5 agreed the architecture transfers to other domains:

\textit{"I could see this working for health behavior change in physical activity or medication adherence."}, P4.

 Qualitative discussions confirmed questionnaire findings. Architectural separation was praised. Explainability emerged as primary weakness, with calls for richer provenance tracking. LLM non-determinism identified as operational risk requiring mitigation. Transferability was also strongly supported.

\begin{tcolorbox}[colback=gray!10,boxrule=0.5pt,title=RQ2 Answer]
We designed an architecture systematically integrating behavioral characteristics through three sequential processing layers and two cross-cutting modules. Ethical compliance is enforced structurally through side-mounted Evaluation implementing Fairness Monitor, Ethics Compliance, and Explainability as first-class architectural concerns. Architect validation confirms successful requirements satisfaction (M=4.62/5), with UI Adaptation highest rated (4.85) and Explainability identified for improvement (4.38). All architects assessed the architecture as transferable across domains. The architecture demonstrates behavioral theory integration and ethical compliance achievable through architectural patterns enforcing constraints at design time.
\end{tcolorbox}

\section{Implementation and User Evaluation}
\label{sec:prototype}

To demonstrate architectural feasibility, we implemented a proof-of-concept in residential energy sustainability. This section describes implementation details and empirical user evaluation.

\subsection{Implementation Context}

Household energy consumption accounts for approximately 26\% of EU greenhouse gas emissions~\cite{consumption}, making residential energy reduction central to sustainability. The domain requires systems adapting interventions to behavioral stages while managing high-consumption appliances.

For the implementation we created a Python backend orchestrating five architectural modules using LLMs for cognitive reasoning via OpenAI API, and React/TypeScript frontend realizing the Adaptation module. Data interchange uses JSON contracts. Code available in \textbf{supplemental materials}.

\subsection{Implementation Walkthrough}

{\em i)} Context Tracker captures device type (desktop vs. mobile) and temporal attributes (time-of-day: morning/afternoon/evening) from session timestamps.

{\em ii)} Behavioral Collector aggregates interaction signals from frontend: hesitation time, clicks, energy consumption, usage hours, user emotions (facial expressions via FaceAPI~\cite{faceapi}). Signals normalized to JSON serving as cognitive profiling input.

{\em iii)} Cognitive Mode uses LLM classifier with Context Tracker + Behavioral Collector data. Specialized prompt analyzes interaction velocity; high clicks + long hesitation on high-wattage appliances indicates analytical processing.

{\em iv)} Behavioral Stage uses LLM classifier mapping users to Transtheoretical Model stages.

{\em v)} Attention Capacity uses LLM classifier estimating cognitive capacity from contextual factors, influencing UI and message complexity.

{\em vi)} Strategy Optimizer uses LLM with Cognitive Mode + Behavioral Stage + Attention Capacity data, selecting optimal intervention from 23-strategy taxonomy~\cite{P1} via constraint-satisfaction prompting.

{\em vii)} Nudge Generator uses selected strategy + appliance data to create message via LLM, enforcing regulatory constraints by integrating Ethics and Fairness prompts. Ensures messages are concise, personalized with consumption data, aligned with user state.

{\em viii)} UI Adaptation uses LLM processing Cognitive Mode + Attention Capacity + Behavioral Stage to dynamically adjust: font size (12px--20px), primary/secondary colors, chart type (bar/pie/line). Ensures UI aligns with cognitive state.

{\em ix)} Explainability generates natural language "Transparency Explanation" revealing system's decision logic, satisfying transparency requirements. Explainability panel screenshot available in the \textbf{supplemental materials}.

{\em x)} Adaptive Dashboard provides interactive environment for monitoring/modifying appliance states. Dynamically renders content based on UI Adaptation output, displays nudges and explanations, connects to feedback system. Adaptive dashboard screenshot is available in the \textbf{supplemental materials}.

{\em xi)} User Feedback captures "thumbs up/down" feedback. Previous nudges tracked via history, informing future Strategy Optimizer decisions.

{\em xii)} Fairness Monitor injects "Bias Mitigation" system prompt into every LLM call, instructing non-discrimination and vulnerable user protection.

{\em xiii)} Ethics Compliance enforces EU AI Act and GDPR constraints by injecting mandatory instructions into LLM prompts as guardrail against manipulative strategies.

{\em xiv)} Outcome Tracker records every intermediate step (raw signals → profiling → strategy → message) and emotional states (pre-nudge and post-nudge) using facial emotion detection. Stored in CSV enabling longitudinal analysis.

\subsection{User Evaluation}
\label{sec:user}

We conducted empirical evaluation with 15 participants interacting with the implemented dashboard to assess feasibility and effectiveness.

We recruited participants through university networks (N = 8) and local community channels (N = 7). Criteria: {\em i)} age 18+, {\em ii)} responsible for household energy decisions. Table~\ref{tab:userdemographics} presents demographics.

\begin{table}[h]
\caption{User Study Participant Demographics (N=15)}
\label{tab:userdemographics}
\centering
\footnotesize
\begin{tabular}{llc}
\toprule
\textbf{Characteristic} & \textbf{Category} & \textbf{N (\%)} \\
\midrule
Age & 18--24 & 5 (33.3\%) \\
 & 25--34 & 6 (40.0\%) \\
 & 35--44 & 3 (20.0\%) \\
 & 45+ & 1 (6.7\%) \\
\midrule
Gender & Male & 8 (53.3\%) \\
 & Female & 7 (46.7\%) \\
\midrule
Education & High school & 2 (13.3\%) \\
 & Bachelor's & 7 (46.7\%) \\
 & Master's/PhD & 6 (40.0\%) \\
\midrule
Energy Management & High awareness & 4 (26.7\%) \\
Experience & Medium awareness & 7 (46.7\%) \\
 & Low awareness & 4 (26.7\%) \\
\midrule
Tech & Smart home user & 6 (40.0\%) \\
Familiarity & Basic tech user & 7 (46.7\%) \\
 & Low tech user & 2 (13.3\%) \\
\bottomrule
\end{tabular}
\end{table}

Sample represents diverse backgrounds. Age distribution skews younger (73.3\% under 35) due to university recruitment. 

Users accessed web dashboard and explored freely. They could add/remove appliances, modify device states, adjust usage hours. Dashboard captured behavioral signals (clicks, hesitation). System monitored emotions via facial recognition throughout session.

After each generated nudge, users read message and explanation, then rated on a Likert scale: {\em i)} Nudge Quality, "This nudge is helpful and appropriate for my current situation"; {\em ii)} Explanation Quality, "The explanation clearly describes why this nudge was shown". Sessions lasted 20-30 minutes. All provided explicit consent.

Our evaluation addresses RQ3 through three measures:

1. Feasibility (Architecture Instantiation): Demonstrated by showing architecture executes across all components with real users. Feasibility established if: {\em i)} User Modeling produces classifications for all sessions, {\em ii)} Strategy Optimizer selects strategies from taxonomy, {\em iii)} Nudge Generator produces coherent messages, {\em iv)} UI Adaptation applies interface changes, {\em v)} Ethics Compliance doesn't block compliant nudges, {\em vi)} system completes full cycle without failures. Reported in Table~\ref{tab:userresults}.

2. Effectiveness (Intervention Quality): Measured through perceived effectiveness via subjective ratings: {\em i)}Nudge Quality, {\em ii)} Explanation Quality. Results in Table~\ref{tab:evaluation}.

3. Emotional Impact (User Experience): Assessed whether adaptive nudges produce positive or negative emotional responses using facial expression analysis. Addresses key risk: personalized interventions might feel invasive, triggering negative affect. Conversely, helpful nudges should produce positive affect.

Facial Expression Analysis: Used FaceAPI~\cite{faceapi}, a JavaScript library implementing Ekman facial emotions~\cite{ekman1992argument}, detecting seven basic emotions: happiness, sadness, anger, fear, disgust, surprise, neutral. Library outputs probability distributions over emotions per frame.

 We focus on happiness because positive-valence emotion indicates receptiveness and sustained positive affect predicts behavior change adherence~\cite{fredrickson2001role}. We report raw probability values from FaceAPI (range: 0.0--1.0, where 1.0 = 100\% confidence).

 Table~\ref{tab:userresults} shows user testing generated 15 complete executions capturing adaptive behavior. System classified majority (73.3\%) as intuitive mode, 26.7\% analytical. Behavioral stage: 40\% pre-contemplation, 26.7\% contemplation, 33.3\% action. Attention capacity: 66.7\% high, 20\% medium, 13.3\% low.

Strategy Optimizer selected interventions based on profiles. Just-in-time prompts emerged most frequently (53.3\%), with remaining 46.7\% across: remind consequences, raise visibility, enable comparisons, reduce distance. UI Adaptation adjusted parameters: font size 16px for 73.3\%, 19px for 13.3\%, 24px for 13.3\%, demonstrating response to attention capacity.

\begin{table}[h]
\caption{User Testing System Classifications (N=15)}
\label{tab:userresults}
\centering
\footnotesize
\begin{tabular}{lllll}
\toprule
\textbf{ID} & \textbf{Cognitive} & \textbf{Stage} & \textbf{Attn} & \textbf{Strategy} \\
\midrule
1 & Analytical & Contemplation & High & Just-in-time \\
2 & Intuitive & Pre-contempl. & High & Remind conseq. \\
3 & Intuitive & Contemplation & High & Enable compar. \\
4 & Intuitive & Action & High & Raise visibility \\
5 & Intuitive & Contemplation & High & Just-in-time \\
6 & Intuitive & Action & Medium & Just-in-time \\
7 & Intuitive & Pre-contempl. & High & Remind conseq. \\
8 & Intuitive & Pre-contempl. & Low & Just-in-time \\
9 & Intuitive & Pre-contempl. & Low & Reduce distance \\
10 & Intuitive & Action & Medium & Just-in-time \\
11 & Intuitive & Contemplation & High & Just-in-time \\
12 & Analytical & Pre-contempl. & Medium & Raise visibility \\
13 & Intuitive & Pre-contempl. & High & Just-in-time \\
14 & Analytical & Action & High & Just-in-time \\
15 & Analytical & Action & High & Raise visibility \\
\bottomrule
\end{tabular}
\end{table}

Table~\ref{tab:evaluation} presents evaluation metrics. Participants rated nudge quality highly (M=4.73, SD=0.46): 73.3\% provided maximum rating (5), 26.7\% rated 4. Explanation quality received slightly lower but positive ratings (M=4.27, SD=0.70): 40\% rated 5, 46.7\% rated 4, 13.3\% rated 3. This suggests while users found nudges highly appropriate, transparency could be strengthened, consistent with architect evaluation.

\begin{table}[h]
\caption{User Evaluation Results}
\label{tab:evaluation}
\centering
\footnotesize
\begin{tabular}{lc}
\toprule
\textbf{Metric} & \textbf{Mean Value} \\
\midrule
Nudge Quality (1--5) & 4.73 (SD=0.46) \\
Explanation Quality (1--5) & 4.27 (SD=0.70) \\
\midrule
Pre-Nudge Happiness & 0.000170 (SD=0.000092) \\
Post-Nudge Happiness & 0.000500 (SD=0.000213) \\
Happiness Increase & 0.000330 (SD=0.000158) \\
\bottomrule
\end{tabular}
\end{table}

Emotional impact analysis revealed measurable positive effects. Pre-nudge happiness scores (M=0.000170) reflected predominantly neutral emotional states. Post-nudge happiness scores (M=0.000500) showed mean increase of 0.000330, with all participants showing positive changes. While absolute values remain small due to FaceAPI probability output range, consistent directional change across all participants indicates adaptive nudge delivery produces positive emotional responses rather than reactance or annoyance.

Complete execution traces demonstrate architecture successfully adapts behavior across multiple dimensions simultaneously. Sessions with low attention capacity consistently received simplified UI configurations, while Strategy Optimizer simultaneously selected low-complexity approaches rather than information-rich strategies. This coordination between User Modeling and both Nudge Intelligence and UI Adaptation validates architectural design's ability to integrate behavioral signals into adaptive responses.

\begin{tcolorbox}[colback=gray!10,boxrule=0.5pt,title=RQ3 Answer]
The instantiated architecture demonstrates effective adaptive behavior in sustainable energy domain. User profiling components successfully classified participants across cognitive mode, behavioral stage, and attention capacity dimensions. System selected contextually appropriate nudging strategies and UI configurations based on classifications. User evaluation confirms high perceived quality for generated nudges (M=4.73/5) and satisfactory explanation quality (M=4.27/5). Emotional impact measurement shows positive affect changes across all participants following nudge delivery, indicating interventions are well-received rather than perceived as manipulative. Feasibility is demonstrated through successful execution of all 15 user sessions without failures, with complete architectural pipeline from behavioral signal capture through adaptive intervention delivery.
\end{tcolorbox}

\section{Threats to Validity}
\label{sec:threats}

\textbf{Internal Validity.} Architects evaluated architecture after detailed explanations, creating priming effects influencing assessments. Future work should employ blind evaluation. User testing lacked baseline measurements and control conditions; observed outcomes may reflect novelty effects. We partially addressed this by capturing pre-nudge and post-nudge emotional states within sessions. Implementation relies on LLMs exhibiting non-deterministic behavior; we mitigated this setting low temperature (0.3) for classification tasks.

\textbf{Construct Validity.} Facial emotion recognition may not validly represent deeper affective states driving sustained behavior change. We triangulated emotion data with explicit user feedback ratings. Future implementations should validate classifications against validated behavioral science measures.

\textbf{External Validity.} Implementation instantiates architecture exclusively in residential energy sustainability. While architects rated it highly transferable, empirical validation across domains (healthcare, finance) remains necessary. Implementation uses specific technologies (React, Python, OpenAI APIs) that may not generalize to resource-constrained environments. Architecture's abstract design remains technology-agnostic, enabling alternative implementations.

\section{Conclusion and Future Work}
\label{sec:conc}

This paper bridges behavioral theory and software architecture by presenting an architecture that translates psychological constructs into executable software components while enforcing ethical compliance structurally. We conducted a literature review identifying 68 nudging strategies, 11 quality attributes, and 3 user profiling dimensions, transformed into architectural requirements with full traceability. The architecture implements three core processing layers (Data Capture, User Modeling, Nudge Intelligence) with two cross-cutting modules (Adaptation, Evaluation) enforcing fairness monitoring and regulatory compliance as architectural guardrails.

Architect validation with 13 professionals confirmed requirements satisfaction and domain transferability. Implementation in residential energy sustainability using LLMs for cognitive reasoning demonstrated feasibility. User evaluation with 15 participants achieved high perceived nudge quality and measurable positive emotional impact. Explicit architectural decisions documented trade-offs between quality attributes, showing how behavioral theory shapes software structure.

\textbf{Future Work.} Three directions emerge: {\em i)} Cross-domain validation, empirically testing architecture in healthcare, finance, e-commerce to strengthen generalizability claims; {\em ii)} Longitudinal studies, tracking users over weeks/months to assess sustained impact and habituation effects; {\em iii)} Enhanced explainability providing richer access to decision states, enabling transparency reports revealing considered alternatives and rejection rationale. 

\section*{acknowledgment}
This research was supported by the Nudge2Green project (1152-00043A), funded by AgriFoodTure, Innovation Fund Denmark.

\bibliographystyle{IEEEtran}
\bibliography{references}
\end{document}